\documentclass{aastex631}

\shorttitle{On the explosive phase of the tearing mode}
\shortauthors{Baty}

\graphicspath{{./}{figures/}}

\usepackage{bm}%
\usepackage{graphicx}%
\usepackage{amsmath}

\def\ltsima{$\; \buildrel < \over \sim \;$}
\def\gtsima{$\; \buildrel > \over \sim \;$}
\def\simlt{\lower.5ex\hbox{\ltsima}}
\def\simgt{\lower.5ex\hbox{\gtsima}}
\DeclareMathOperator{\sech}{sech}

\begin{document}

\title{On the explosive phase of the tearing mode in double current sheet plasmas: effect of the
equilibrium magnetic configuration on the onset threshold and growth rate }

\author[0000-0003-1925-3983]{Hubert Baty}

\affiliation{Observatoire Astronomique de Strasbourg,
Universit\'e de Strasbourg \\
11 Rue de l'universit\'e, 67000 Strasbourg, FRANCE \\
\textcolor{blue} {hubert.baty@unistra.fr}  }



\begin{abstract}
Magnetic reconnection associated with the tearing instability occurring in double-current sheet systems is investigated within the framework
of reduced resistive magnetohydrodynamics (MHD) in a two-dimensional Cartesian geometry.
The explosive non linear phase is particularly explored using the adaptive finite-element FINMHD code.
The critical aspect ratio, that is defined as the minimum $L/x_s$ ratio (with $L$ and $x_s$ being the periodic system length and half-distance
between the two current layers respectively) necessary for non linear destabilization after the linear and early
non linear saturation phases, is obtained. The latter threshold is independent of the details of the chosen initial equilibrium (double Harris-like
magnetic profile) and of the resistivity. Its value is shown to be $4.7$, that is close and slightly smaller than the value of order $5$ deduced using
a more particular equilibrium configuration in previous studies. The time dependence of the kinetic energy ($E_K$) 
is shown to follow a double exponential law, $E_K \propto \exp \ [e^{(\gamma^* t)} ]$, with a pseudo-growth rate $\gamma^* \simeq 0.1 \ t_A^ {-1}$
($t_A$ being the characteristic Alfv\'en time) that is again independent of the configuration and resistivity.
The mechanism offers a possible explanation for the sudden onset of explosive magnetic energy release occurring on the fast Alfv\'en time scale in
disruptive events of astrophysical plasmas with pre-existing double current sheets like in the solar corona.
\end{abstract}



\section{Introduction} \label{sec:intro}

The tearing mode developing in current sheet plasmas is believed to play a significant role during
explosive phenomena observed in different magnetized astrophysical systems like the
solar corona, the solar wind, or pulsar winds. When a single electric current sheet (corresponding to a magnetic reversal) is present, the tearing instability
is considered to be a slow mode as it involves a magnetic reconnection process in the form of magnetic islands growing on a resistive
time scale (except if plasmoid chains are able to develop in cases of very long and thin layers).
The tearing mode of  such highly conducting plasmas under consideration in this study has been extensively investigated in planar and sheet pinch configurations
\citep {pri00}. The latter magnetohydrodynamic (MHD) instability is characterized by a threshold for the wavenumber $k$, as $k \le k_M$ is required.
The maximum wavenumber $k_M$ depends on the details of the equilibrium profile, and its value is $k_M = 1/a$ for the standard
Harris profile, where $a$ defines the half-thickness of the current sheet. This linear stability threshold can be easily translated into a condition
$L \ge L_{min}$ for the length of the system, where $L_{min} = 2 \pi/k_M$ \citep {bisk09}. These resistive instabilities are admitted to
grow on a characteristic time scaling as $S^{1/3}-S^{3/5}$, where $S$ is the Lundquist number defined as $S = a V_A/\eta$ (with $V_A$
and $\eta$ being the Alfv\'en velocity and the small magnetic diffusivity parameter respectively). For example, as the typical $S$ value for the solar coronal plasma is
$\sim 10^{10}$, the characteristic tearing time scale cannot account for fast time scales involved in solar flares by many orders of magnitude
(i.e. a few years/months versus a few minutes).

On another hand, multiple current sheets due to many field reversals are also expected to form in such plasmas,
giving rise to multiple tearing instabilities. A richer dynamics is thus expected 
with a multi-regime feature for the reconnection process and associated magnetic energy release, that is not observed for a single current sheet.
This is the case for a double current sheet system. Interestingly, for this latter configuration, an explosive growth phase can be
obtained in association with the so called double tearing mode (DTM hereafter). Indeed, after a long-time-scale evolution of slow resistive linear growth (similar
to the growth of a single tearing mode with similar scaling laws) followed by a phase of saturation, the DTM suddenly changes to rapid growth \citep {ish02, wan07, jan11, zha11, akr17}.
During this explosive phase, the kinetic energy associated to the flow can increase by several orders of magnitude in a few Alfv\'en times quasi-independently of the resistivity $\eta$ or Lundquist number $S$.
Extensive studies using the simplest MHD framework (i.e. reduced MHD equations in a two-dimensional planar geometry) have shown that the explosive
phase is triggered by a structure-driven non linear instability \citep {ish02, jan11}. The onset of this ideal MHD instability is ascribed to a critical deformation of
the magnetic islands growing on the two current layers. The explosive reconnection is triggered only when the aspect ratio $L/x_s$ 
(with $L$ and $x_s$ being the periodic system length and the half-distance between the two current layers respectively) is
higher than a critical value, that is $\simeq 5$, independently of the resistvity. Otherwise the system with the two magnetic islands no longer evolves. This second critical
length value $L_c$ is in general higher than the $L_{min}$ one.

However, the previously cited studies have mostly employed a particular magnetic field profile for the initially unstable equilibrium, which is
generally assumed in the context of laboratory plasmas. The scope of the present study is precisely to test the robustness of this critical aspect ratio value
by employing another profile which is the generalization of a Harris-type profile generally used for tearing mode of a single current sheet.
Moreover, our profile parametrization allows to study the effect of the shear parameter (deduced from the thickness of the two current
layers at the reversal location of the magnetic field), independently of the distance between the two layers $2 x_s$.
Finally, we investigate how the fast Alfv\'enic time scale (previously reported) for the non linear growth
depends on the initial equilibrium and also from the 'distance' to the non linear threshold. We use a strongly adaptive finite-element code, FINMHD, that has been
specifically designed to address such reconnection problem within the framework
of reduced resistive MHD in a two-dimensional Cartesian geometry.
The outline of the paper is as follows. In Section 2, we briefly present the code and the initial
setup. Section 3 is devoted to the presentation of the results. Finally, we conclude in section 4.

\section{FINMHD code and initial setup}
The usual set of reduced MHD equations in two dimensions (2D) (i.e. $x-y$ plane) is considered to be a good approximation to represent the dynamics in a plane perpendicular to a dominant 
and constant magnetic field component ($B_z$). In this way, the incompressibility assumption in the 2D plane is admitted to be well justified. Instead of taking a standard formulation in terms of
perpendicular velocity and magnetic vectors ($\bm V$ and $\bm B$ respectively), it is generally advantageous to use another formulation with scalar variables like stream functions (hereafter 
$\phi$, and $\psi$), as this ensures the divergence-free property for these two vectors. Moreover, in order to facilitate the numerical implementation,
a (dimensionless) model using the electric current density $J$ and the flow vorticity $\omega$ for the main variables is adopted in FINMHD \citep {bat19},
\begin{equation}  
      \frac{\partial \omega}{\partial t} + (\bm{V}\cdot\bm{\nabla})\omega = (\bm{B}\cdot\bm{\nabla})J + \nu \bm{\nabla}^2 \omega ,
\end{equation}
\begin{equation}      
        \frac{\partial J }{\partial t} + (\bm{V}\cdot\bm{\nabla})J =  (\bm{B}\cdot\bm{\nabla})\omega + \eta \bm{\nabla}^2 (J -J_e) +  g(\phi,\psi) ,
\end{equation}
\begin{equation}                 
     \bm{\nabla}^2\phi = - \omega ,
 \end{equation}
\begin{equation}                        
     \bm{\nabla}^2\psi = - J ,  
\end{equation}
with $g(\phi,\psi)=2 \left[\frac{\partial^2 \phi}{\partial x\partial y}\left(\frac{\partial^2 \psi}{\partial x^2} - \frac{\partial^2 \psi}{\partial y^2}\right) - \frac{\partial^2 \psi}{\partial x\partial y}\left(\frac{\partial^2 \phi}{\partial x^2} - \frac{\partial^2 \phi}{\partial y^2}\right)\right]$. We
have introduced the two stream functions, $\phi (x, y)$ and $\psi (x, y)$, defined as $\bm{V} = {\nabla} \phi \wedge \bm{e_z}$ and $\bm{B} = {\nabla} \psi \wedge \bm{e_z}$ ($\bm{e_z}$
being the unit vector perpendicular to the $xOy$ simulation plane).
Note that $J$ and $\omega$ are the $z$ components of the current density and vorticity vectors, as $\bm{J} = \nabla \wedge \bm{B}$ and $\bm{ \omega} = \nabla \wedge \bm{V}$ respectively (with units using $\mu_0 = 1$). Note also that we consider the resistive diffusion via the $\eta \bm{\nabla}^2 J $ term ($\eta$ being the resistivity assumed uniform for simplicity), and also a viscous term
$\nu \bm{\nabla}^2 \omega$ in a similar way (with $\nu$ being the viscosity parameter).
The above definitions results from the choice $\psi \equiv A_z$, where $A_z$ is the $z$ component of the potentiel vector $\bm{A}$ (as $\bm{B} = \nabla \wedge \bm{A}$).
FINMHD code is based on a finite element method using triangles with quadratic basis functions on an unstructured
grid. A characteristic-Galerkin scheme is chosen in order to discretize in a stable way the Lagrangian derivatives
appearing in the two first equation. Moreover, a highly adaptive (in space and time) scheme
is developed in order to follow the rapid evolution of the solution, using either a first-order time integrator (linearly unconditionally stable) or a second-order one (subject to a CFL time-step restriction). Typically, a new adapted grid can be computed at each time step, by searching the grid that renders an estimated error nearly uniform. More precisely, the method allows to always cover the current structures with
a few tens of triangles at any time, by using the Hessian matrix of the current density as the main refinement parameter.
The technique used in FINMHD has been tested on challenging tests, involving unsteady strongly anisotropic solution for the advection equation, formation of shock structures for viscous Burgers equation, and magnetic reconnection for the reduced set of MHD equations. 
The reader should refer to \cite {bat19}
for more details on the numerical scheme and also to the following references for applications to different aspects of
magnetic reconnection in MHD framework  \citep {bat20a, bat20b, bat20c}.

\begin{figure}
\centering
 \includegraphics[scale=0.54]{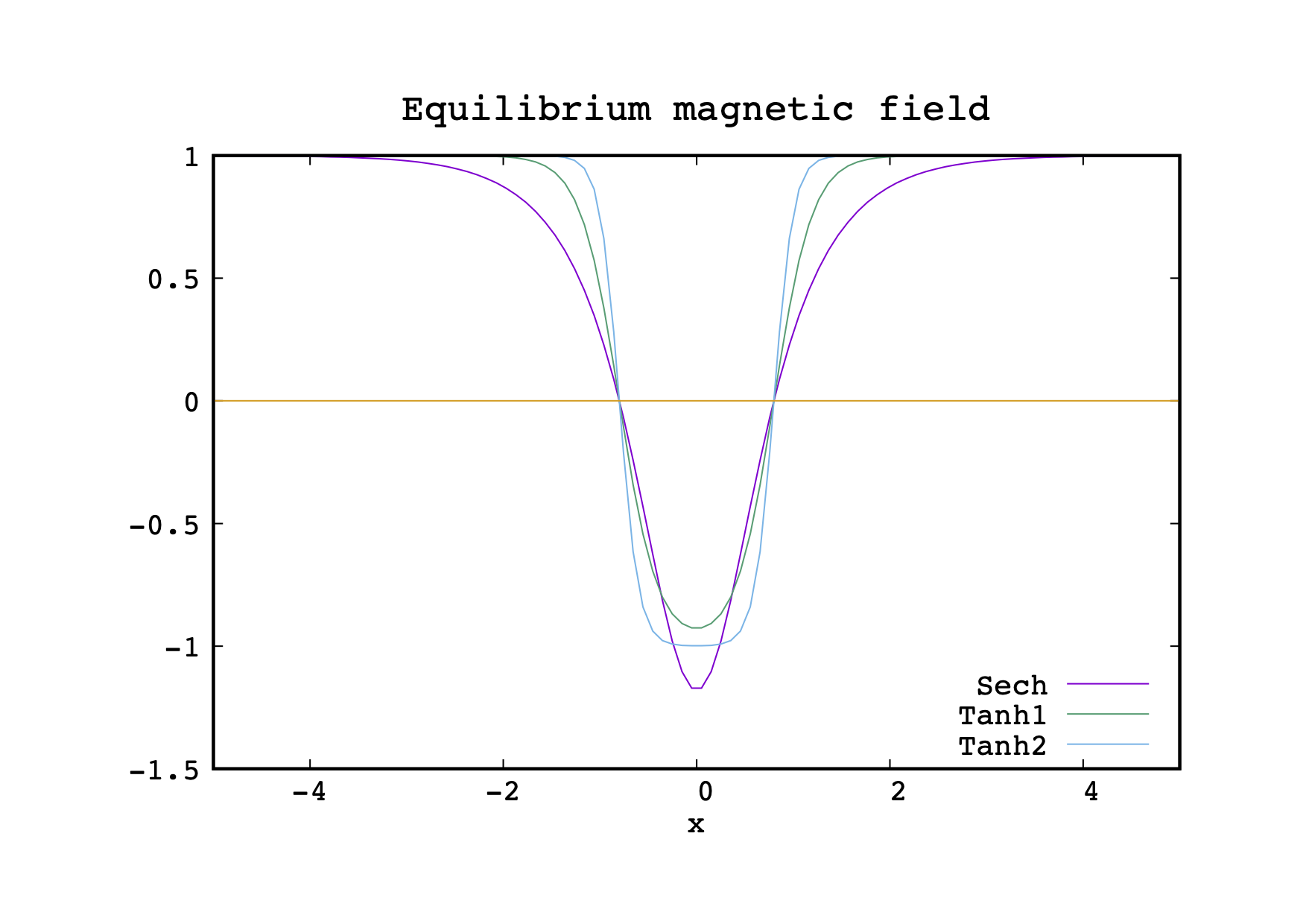}
  \caption{Initial configuration for the equilibrium magnetic field $B_e (x) /B_0$ for the 'sech' profile and for two Harris-like 'tanh' profiles
  (using $a = 0.2$ for 'tanh2', and $a = 0.4$ for 'tanh1'). The field reversals (intersection with horizontal line) are situated at $x = \pm x_s = \pm 0.8$ for the different equilibria.
  }
\label{fig1}
\end{figure}

An initial one dimensional equilibrium is considered by taking an analytical expression for the $y$ component of the magnetic field profile $B_y = B_e (x)$. In the present work,
a double Harris-like profile (quoted as 'tanh' below) is assumed,
\begin{equation}  
      B_e (x) = B_0 \left [ 1 + \tanh  \left (  \frac{x - x_s }{a} \right )  -  \tanh  \left (  \frac{x + x_s }{a}   \right ) \right ] ,
\end{equation}
with two hyperbolic tangent-like reversals at $x = \pm x_s$, a magnetic shear defined by the length scale $a$, and an asymptotic magnetic field amplitude $B_0$ at large distance.
 The corresponding current density is,
 \begin{equation}  
      J_e (x) =   \frac{B_0 }{a} \left [    \frac{1 }{ \cosh^2  [  (x - x_s)/a  ]  }   -  \frac{1 }{ \cosh^2  [  (x + x_s)/a  ]  }   \right ]  ,
\end{equation}
which is also used in Equation (2) in order prevent the diffusion of the ideal equilibrium via the $- \eta \bm{\nabla}^2 J_e $ source term. Indeed, such diffusion is unwanted
when the resistivity parameter is not small enough. Moreover, a true ideal MHD
equilibrium should require a thermal pressure term $P(x)$ or a $x$ dependence for a perpendicular magnetic field component (i.e. $B_z (x)$ not included in our model) in order to ensure the equality,
$\bm{J}  \times \bm{B} -  \nabla P  = 0 $ in the momentum MHD equation. However, working with the vorticity equation only requires $\bm{B_e}\cdot\bm{\nabla}J_e$
to vanish, which is automatically satisfied (see Equation 1).

In most previous studies (see \cite {jan11} for example), another equilibrium magnetic field configuration (quoted as 'sech' below) was applied as,
\begin{equation}  
      B_e (x) = B_0 \left [ 1 - (1 + b_c)  \sech (\zeta x)    \right ] ,
\end{equation}
where the parameters $b_c$ and $\zeta$ are chosen in such a way to set the reversal locations at $x_s$ via $ \sech  (\zeta x_s)  = 1 / (1 + b_c)$,
the local magnetic shear being equal to $\pi/2$. For example, values of $b_c = 1.18$ and $\zeta = 1.77$ were used for an equilibrium with
$x_s = 0.8$, which is plotted in Figure 1. One can also compare the latter profile with our 'tanh' equilibria using two different magnetic shear values, i.e.
for current layer half-thickness $a = 0.2$ and $a = 0.4$.

\section{Results}
In order to have an overview of the scenario leading to the explosive phase during the DTM evolution, we present below the results
of a representative case of an unstable 'tanh' equilibrium profile with $x_s = 0.8$ and $a = 0.3$. The periodic longitudinal length value is chosen to be $L = 4$. In this case,
one can first check that DTM mode is linearly unstable as the corresponding maximum normalized number $k a= 2\pi a/L \simeq 0.47 $ is indeed lower than unity. Equivalently, the length value
$L$ is higher than the critical minimum value, $L_{min}$, for linear destabilization of the DTM for the chosen 'tanh' equilibrium profile, that is $L_{min} \simeq 1.88$.
The chosen resistivity parameter is $\eta = 10^{-3}$, and the magnetic Prandtl number is $P_r = \nu / \eta = 1/3$. In this work, we consider the low
viscosity regime in order to compare to previous works where a zero viscosity was adopted \citep {jan11}.

Note also that, we choose
$B_0 = 1$ defining thus our normalization. Consequently, the time variable $t$ is normalized using the Alfv\'en time $t_A = l /V_A$,
where $l$ is the unit distance and $V_A$ is the Alfv\'en velocity based on $B_0$ (i.e. $V_A = 1$).
In this work, the simulation domain is situated in the $x$-range $[- L_x/2 : L_x/2]$ with $L_x = 4$ corresponding to an outer boundary
placed sufficiently far enough away in order to not influence the main dynamics. Fixed boundary conditions are imposed
at $x = \pm L_x/2$. Periodic boundary conditions are assumed in
the $y$ direction in order to select different $k_m$ wavenumber values according to $k _m = 2 \pi m / L$ ($m$ being an integer) for a given $L_y = L$ value. The linearly
fastest DTM instability in this work corresponds to a $m = 1$ mode, thus involving a single $m = 1$ magnetic island growing on each current layer (see below).

For this representative case, as a result of the adaptive refinement strategy, the minimum reached edge size of the triangles during the simulation varies between $h_{min} \simeq 0.01$ (necessary to resolve the
equilibrium structure at very early times) and $h_{min} \simeq 0.001$ (necessary to resolve intense localized secondary current layers during strongly non linear phase).
The imposed maximum edge size is $h_{max} \simeq 0.1$ mainly covering regions where the current density is very low. The corresponding total number of triangles varies between
$15000$ and $35000$ approximately. One must note that, a higher number of triangles with a smaller minimum edge size is consequently required when a smaller
resistivity is employed.

\begin{figure}
\centering
 \includegraphics[scale=0.30]{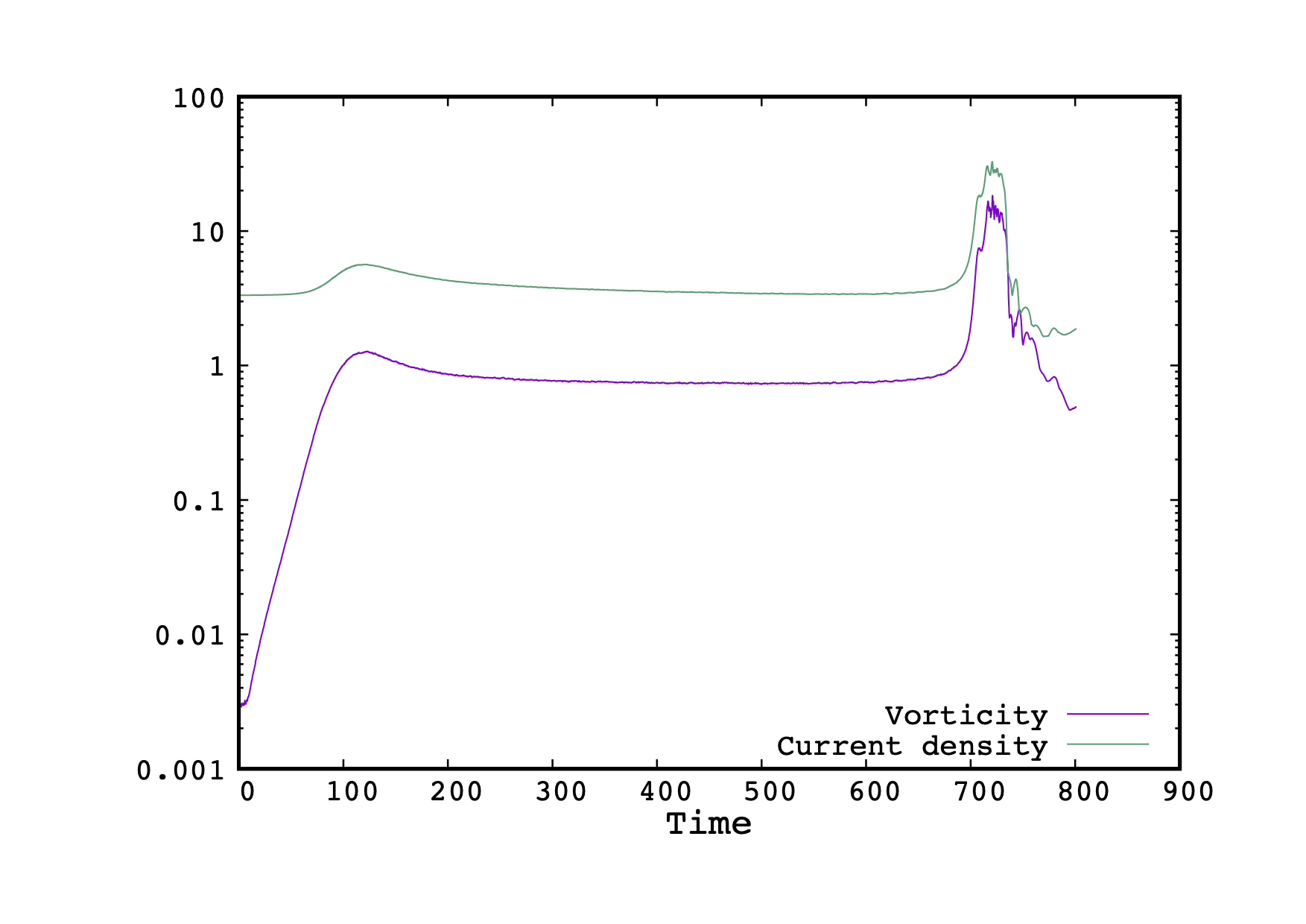}
  \includegraphics[scale=0.30]{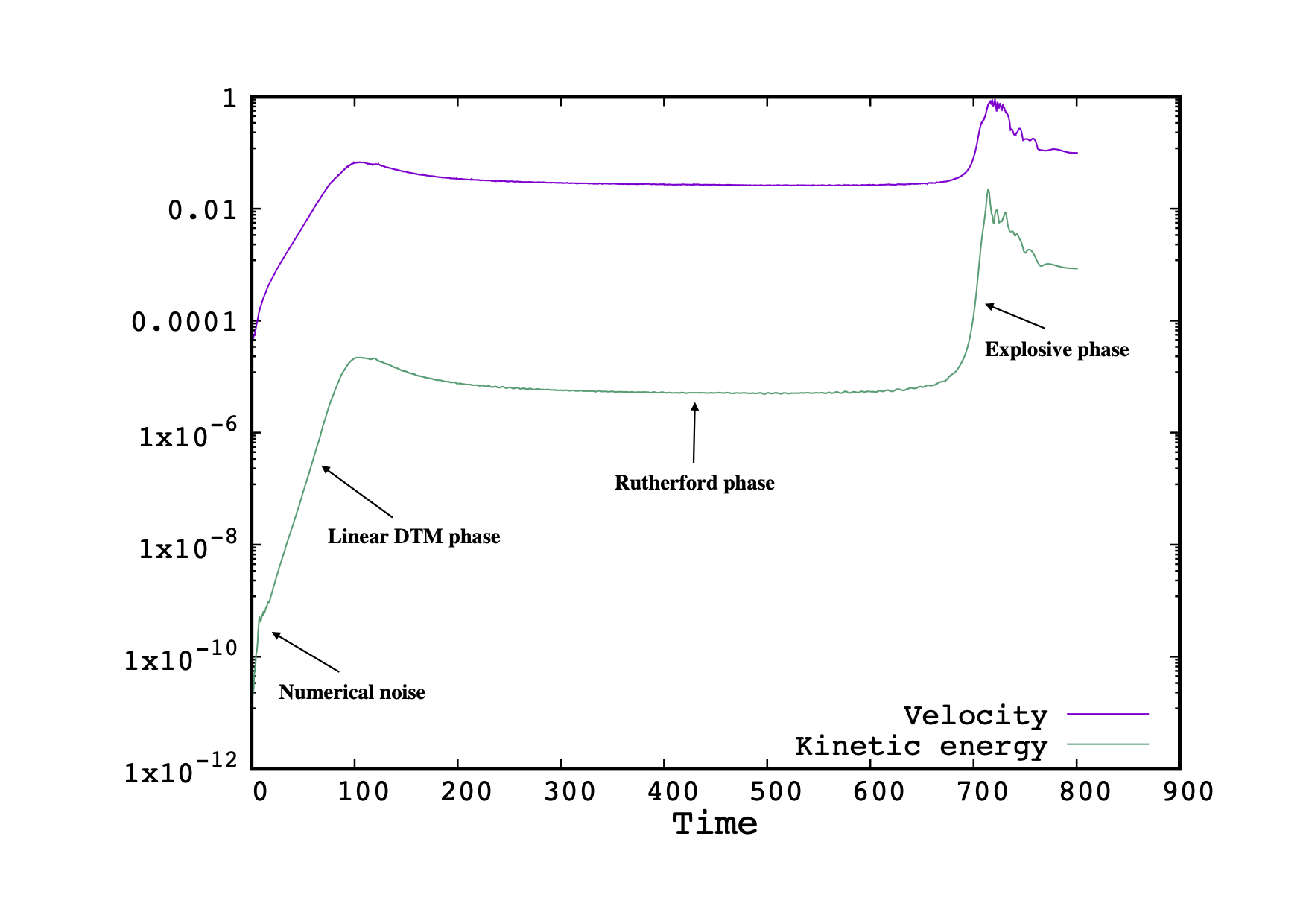}
  \caption{(Left panel) Time evolution of the maximum vorticity $\omega_M$ and associated maximum current density $J_M$ for a case exhibiting an explosive non linear evolution 
  at $t \simeq 700$ , and (right panel) corresponding time evolution of the maximum velocity ($V_y$ component) and kinetic energy $E_k$ (right panel). The time
  is expressed in Alfv\'en time unit $t_A$ (see text).
  }
\label{fig2}
\end{figure}

\begin{figure}
\centering
 \includegraphics[scale=0.185]{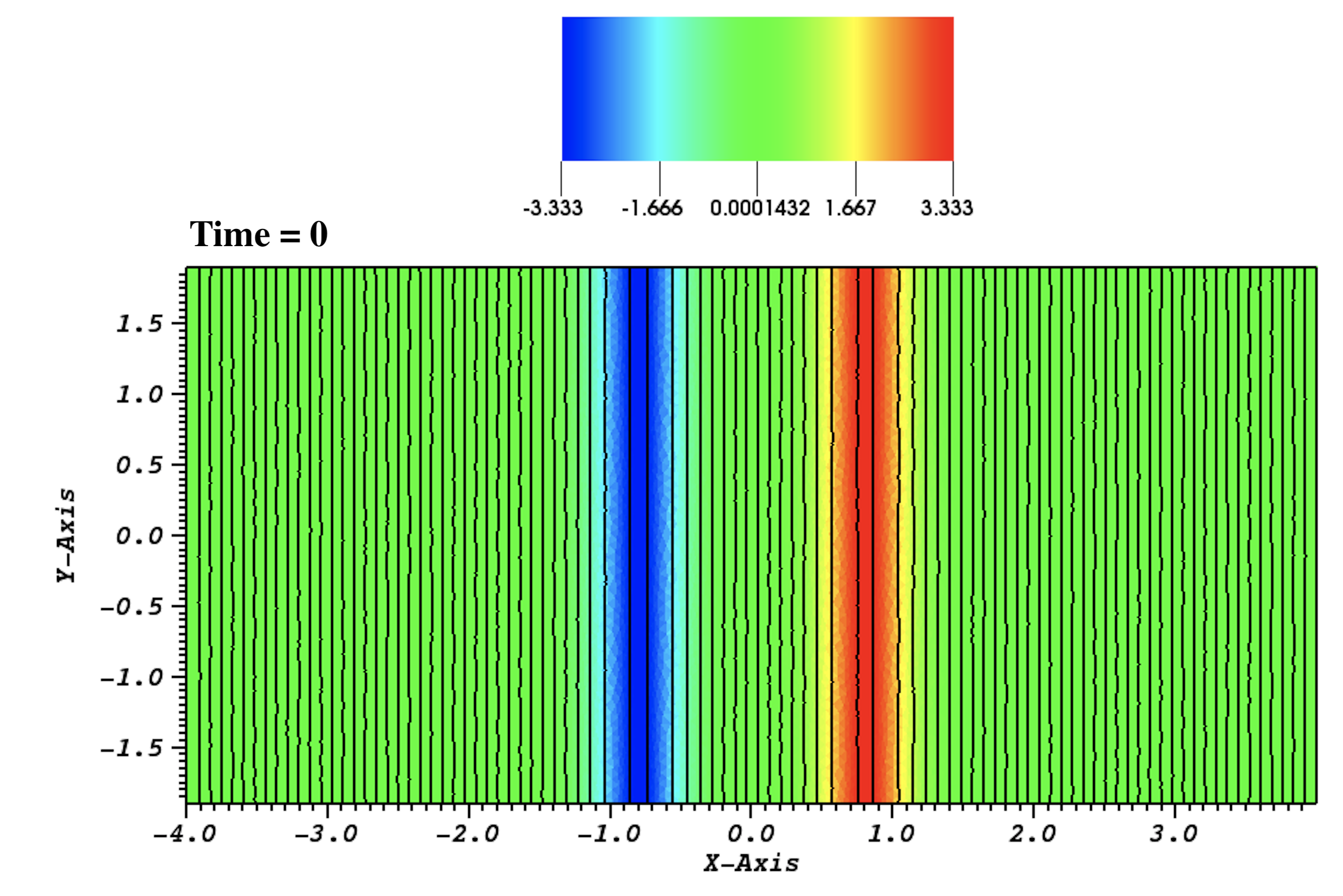}
 \includegraphics[scale=0.185]{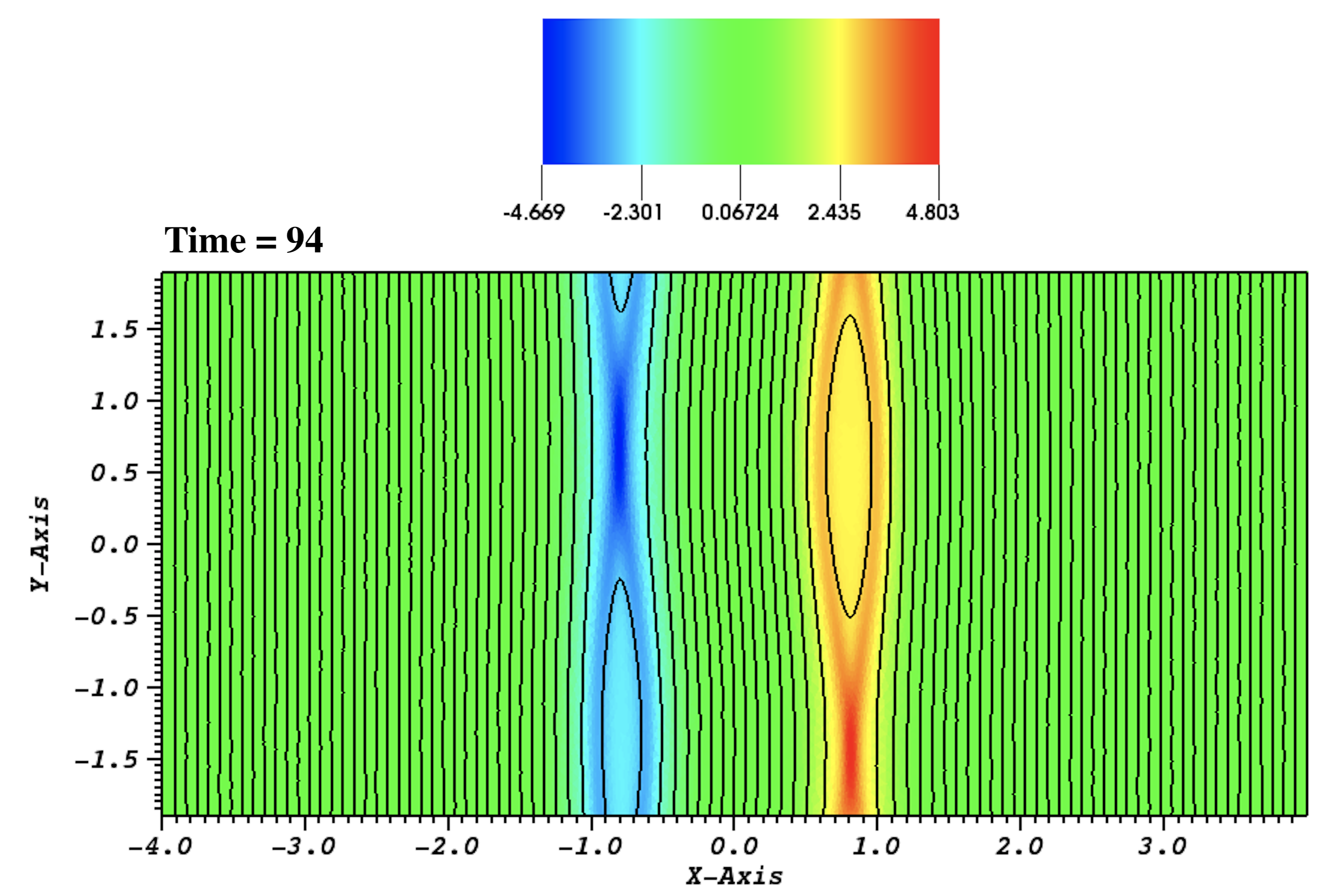}
  \includegraphics[scale=0.185]{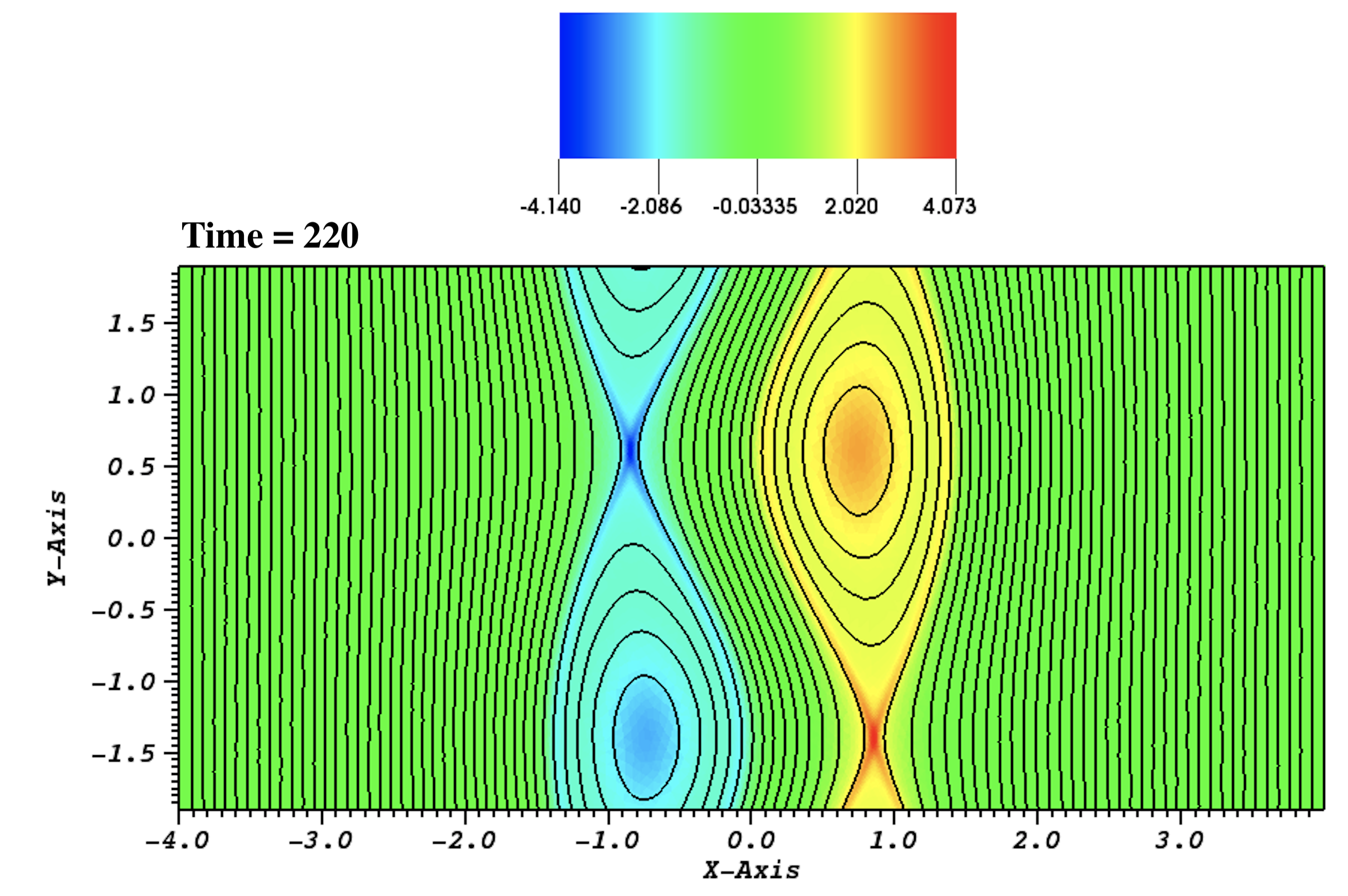}
 \includegraphics[scale=0.185]{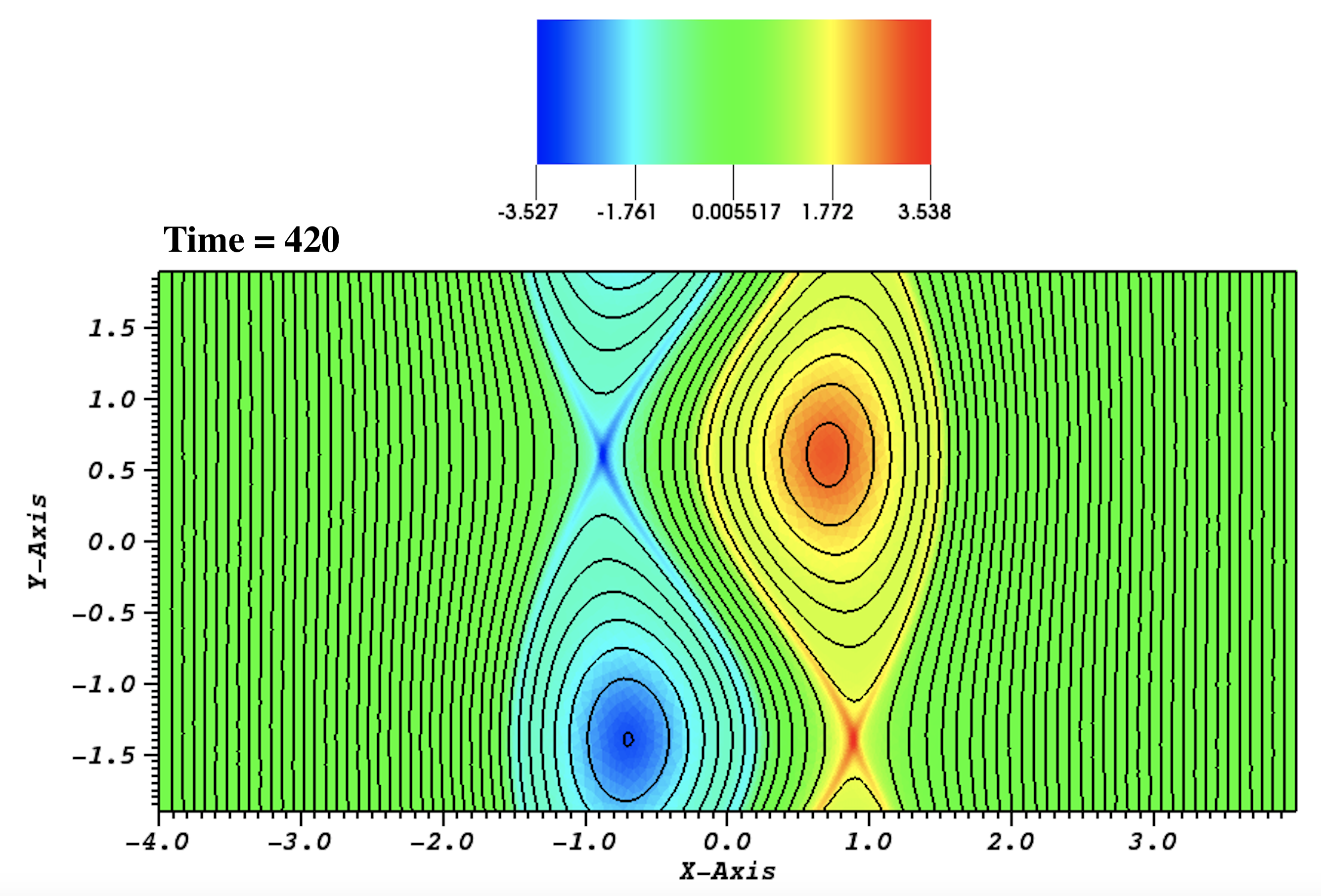}
  \caption{Colored iso-contour map of the current density at different selected times during the representative simulation case of previous figure (see also text),
   overlaid with corresponding selected magnetic field lines. Only early times corresponding to the two first phases are taken.
  }
\label{fig3}
\end{figure}

\begin{figure}
\centering
 \includegraphics[scale=0.185]{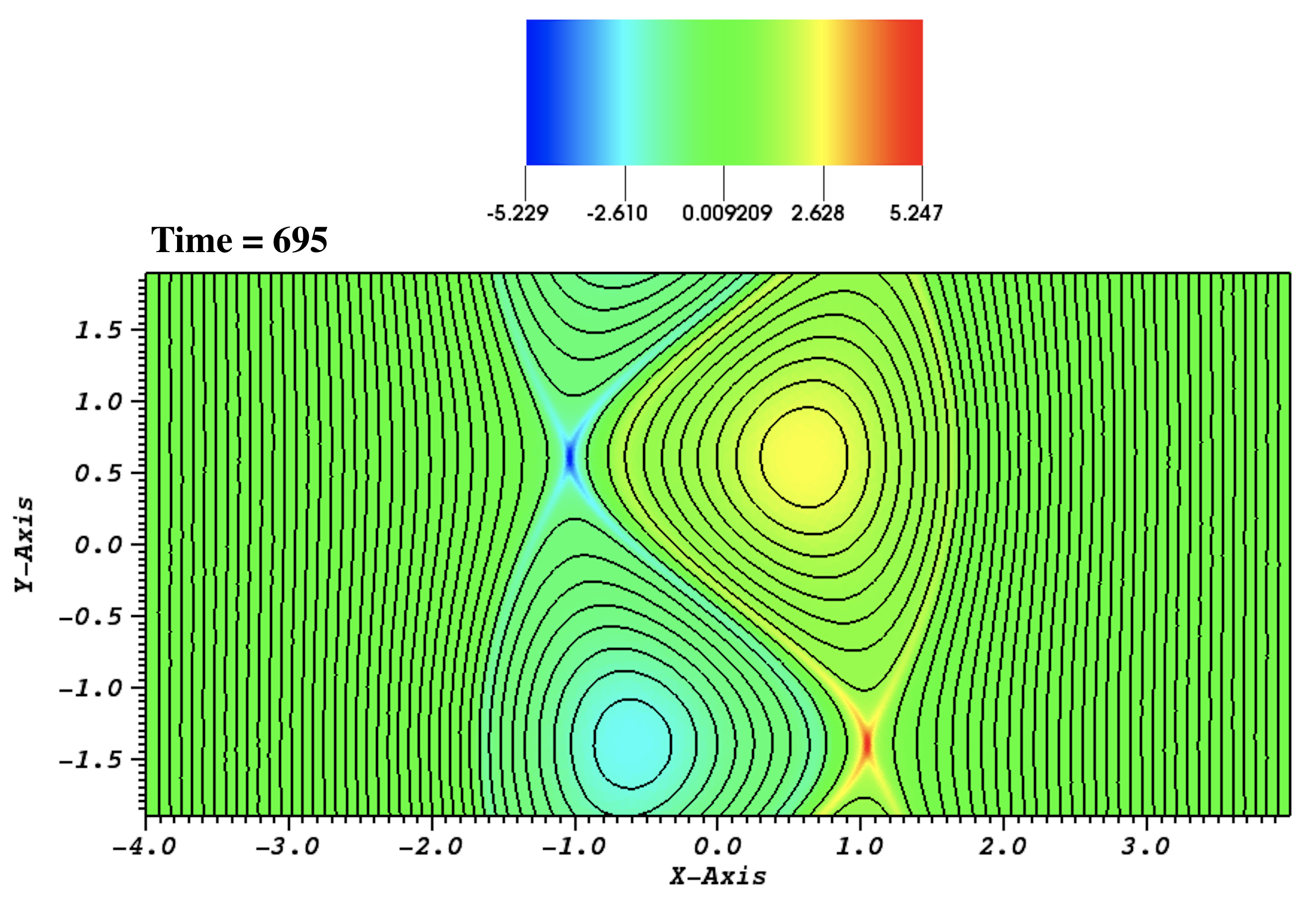}
 \includegraphics[scale=0.185]{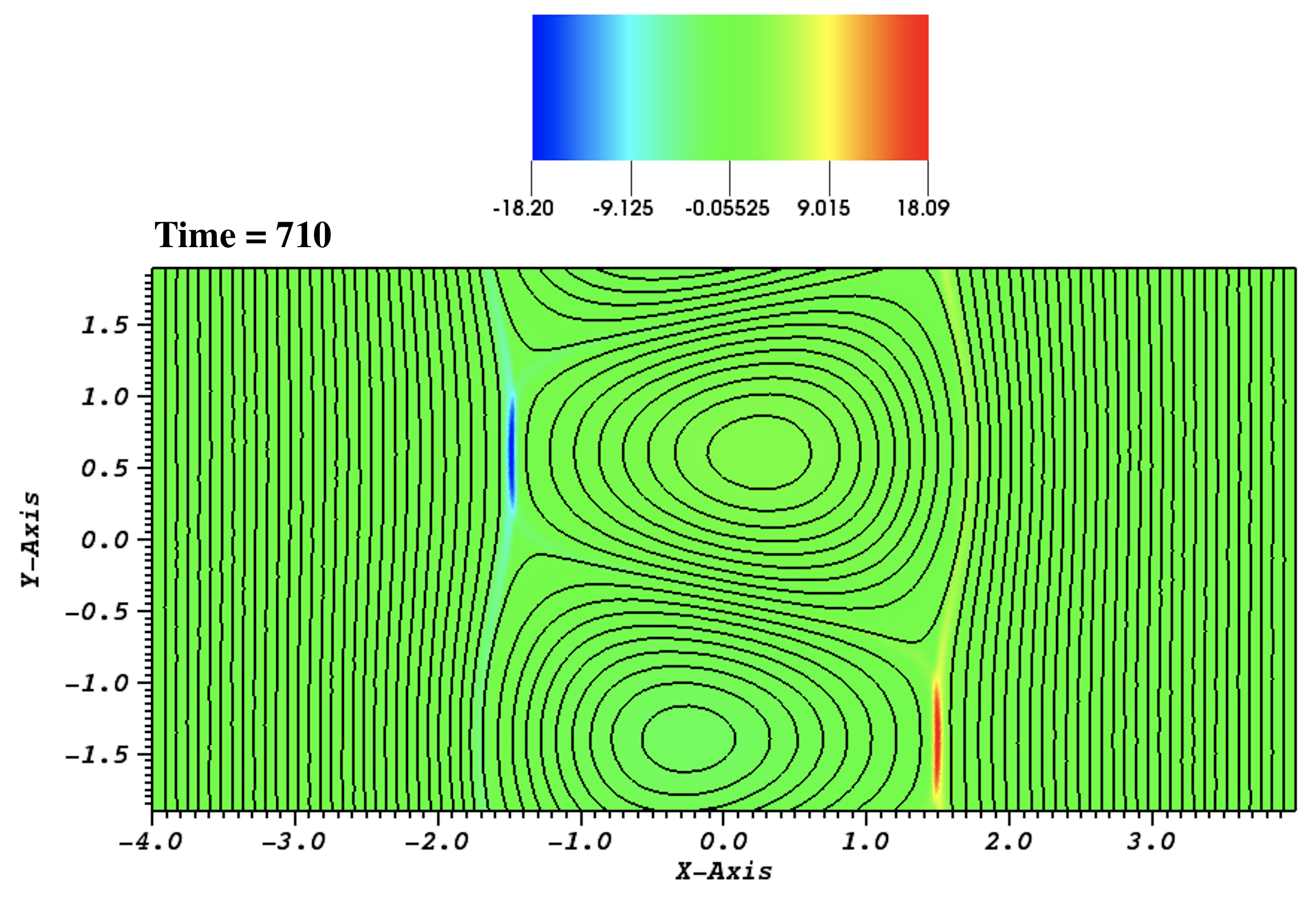}
  \includegraphics[scale=0.185]{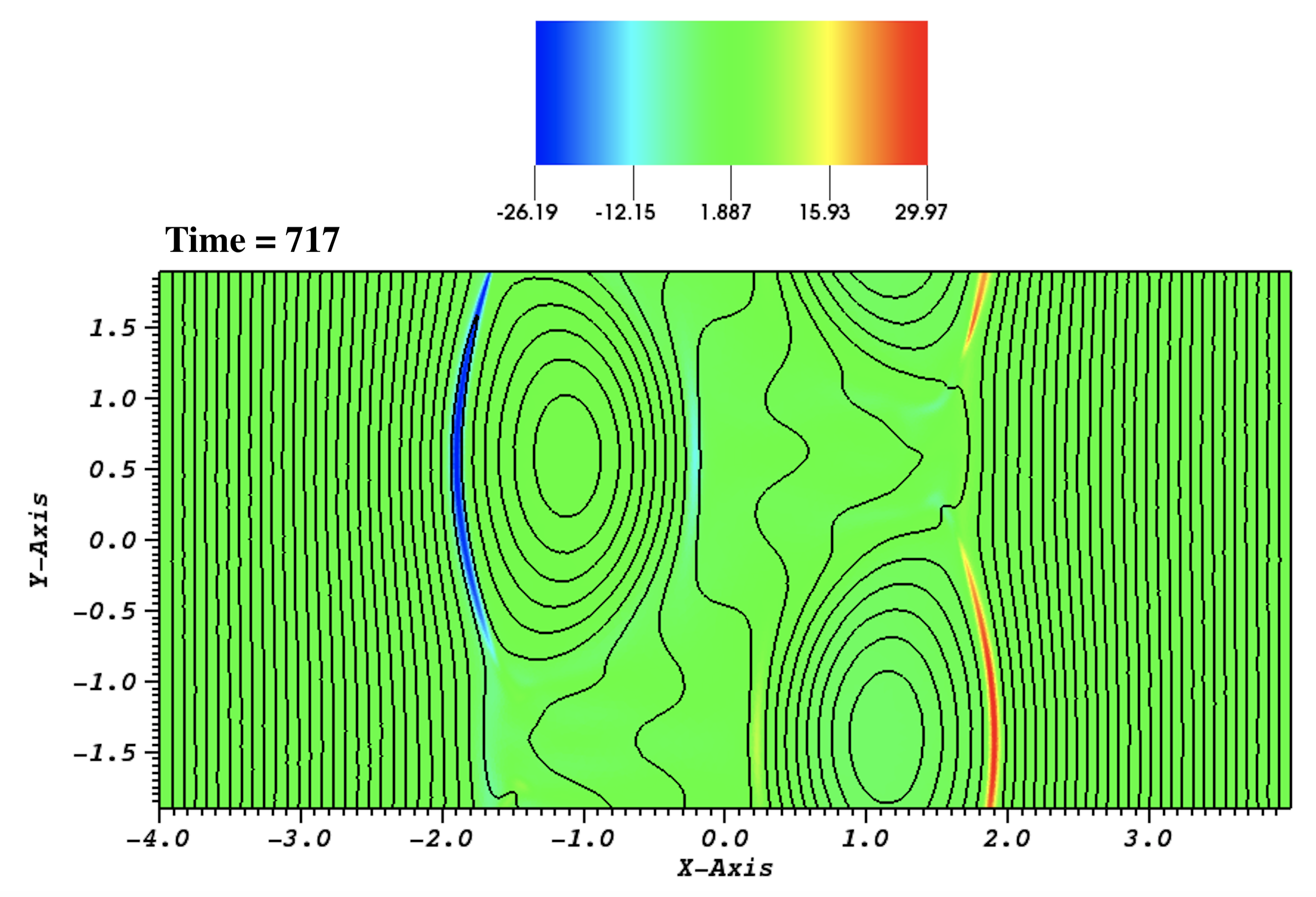}
 \includegraphics[scale=0.185]{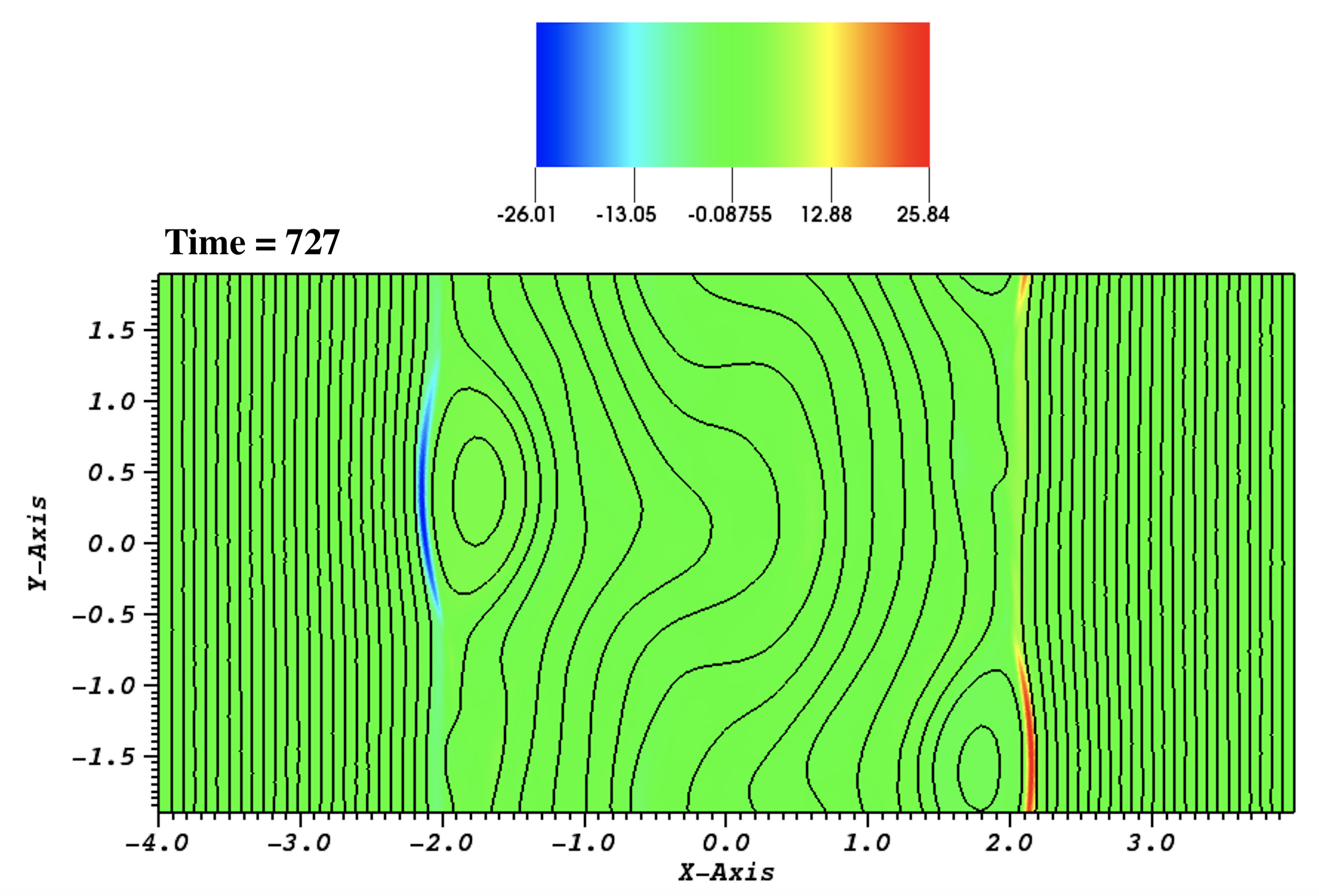}
  \caption{Same as previous figure for later times $t  \simgt 700$ focusing on the explosive phase.}
\label{fig4}
\end{figure}

\subsection{Overview of the explosive DTM dynamics}
Figure 2 (left panel) shows the time evolution of the maximum vorticity $\omega_M$ and of the maximum current density $J_M$ measured over
the whole computational domain, whilst the corresponding maximum velocity amplitude ($V_y$ component) and the integrated kinetic energy $E_K$ are
plotted in right panel. At early time, after a short period of oscillating numerical noise (that is barely visible in $E_K$ curve for example),
one can clearly see for $t \simlt 100$ the linear development of the DTM exhibiting a linear dependence with time (in our semi-log representation) for $\omega_M$, $V_y$, and $E_K$.
The slope measured for $E_K$ is obviously twice the slope for $V_y$ or $\omega_M$, and represents the linear growth rate. More precisely,
 one can deduce the linear growth rate $\gamma_l$ of the DTM instability, as $\gamma_l = \partial_t E_K / E_K$. 
 Note that $\gamma_l \propto \eta^{3/5} \propto S^{-3/5}$ for such resistive instability  \citep {bisk09}.
 Figure 3 illustrates the spatial structure associated to the deformation of the current density and resulting magnetic field topology
 at different times. As expected, the DTM is characterized by two magnetic islands growing on the two initial current layers.
 Moreover, the corresponding observed eigenmode appears to be anti-symmetric, with an out-of-phase island structure between the two
 current layers. Indeed, the $O$-point of the magnetic island at one current sheet is facing with the $X$-point of the island
 at the other current sheet. As shown from linear stability analysis,
 this type of anti-symmetric mode, called A-mode \citep {wei20}, is expected
 to dominate the second type of mode (symmetric one) called S-mode.
 
 The linear phase is followed by a saturation called a Rutherford regime, that has been extensively studied for single tearing modes. In this regime, $E_K$ remains
 quasi-steady, but the perturbed magnetic energy (not shown) continues to grow algebraically (see \cite {jan11} and references therein) as the flow and magnetic
 flux are decoupled. This second phase exhibits a rather long evolution (at least for the parameters chosen in this case) with the two
 islands growing on a pure resistive time scale (see for example the snapshots at $t = 220$ and $t = 420$ in Figure 3).

After this Rutherford phase, an abrupt growth in all the physical variables is observed at $t \sim 700$ in Figure 2. The corresponding contour plot of the current density
overlaid with magnetic field lines (in Figure 4) show that this third phase begins with a sudden triangular deformation of the islands
(snapshot at $t = 695$). This is followed by an increase of the current density at the two $X$-points (see snapshot at $t = 710$),
which drives a second phase of magnetic reconnection, ending when all the closed field lines situated inside the magnetic islands
have reconnected with the external ones (i.e. at $t \simeq 730$). The final state at the end of the simulation tends to be a relaxed configuration
free of magnetic islands and current layers, and the corresponding magnetic field lines tend to become straight again.

Note that, the process of magnetic reconnection during the latter phase is inverse when compared to the one associated to the linear
DTM, as the closed field lines where forming during islands linear growth instead of disappearing during the explosive phase. The mechanism at the origin of this explosive reconnection dynamics was
identified as a structure-driven instability, with a threshold ascribed to a critical magnetic island deformation \citep {jan11}. In summary, it is shown that the nonlinear destabilization is obtained
only when the ratio $L/x_s$ is higher than a value of order $5$ (see Figure 2 in \cite {jan11}) for the 'sech' profile. Here, we have used
a case for the 'tanh' profile having exactly this critical ratio, as $L/x_s = 5$ for this illustrative case.

So, let us now investigate the existence/dependence of this critical aspect ratio value on the initial magnetic equilibrium.

\begin{figure}
\centering
 \includegraphics[scale=0.52]{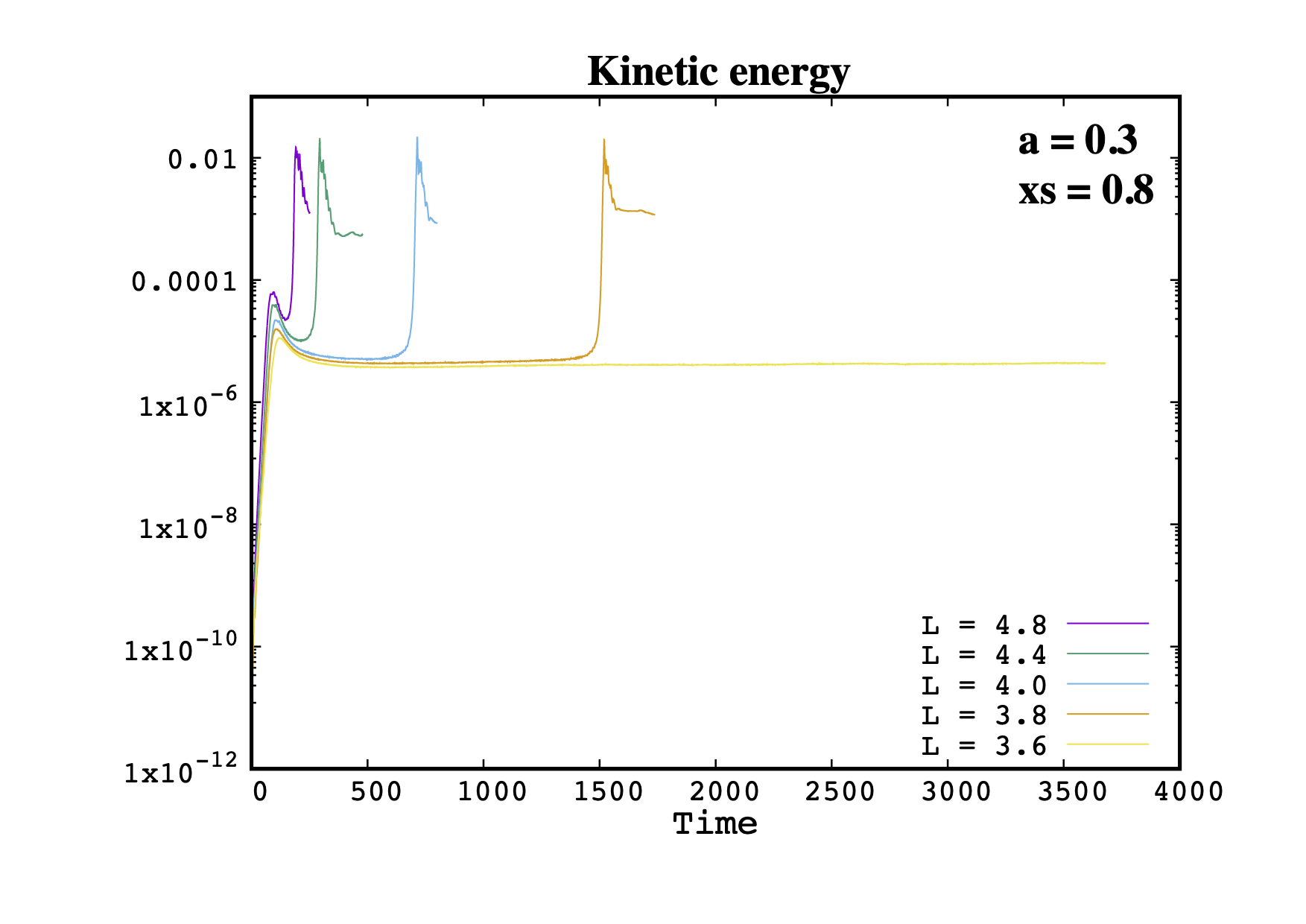}
  \caption{Time evolution of the kinetic energy $E_K$ for different cases employing five different length values $L = 4.8, 4.4, 4.0, 3.8,$ and $3.6$.
  The initial 'tanh' equilibrium with $x_s = 0.8$, and $a = 0.3$, is taken. The other parameters are  $\eta = 10^{-3}$ and $P_r = 1/3$.
   }
\label{fig5}
\end{figure}

\subsection{On the critical aspect ratio for explosive reconnection}

Following the procedure used in \cite {jan11}, we first explore different $L$ cases for a fixed $x_s$ value, that is
$x_s = 0.8$ as taken for the reference case just above. The
results using the kinetic energy evolution as the main diagnostic to discriminate between stability and nonlinear destabilization,
are plotted in Figure 5. Indeed, one can clearly see that the critical $L_c$ value for our configuration is estimated to be close to $3.7$,
as the case using $L = 3.6$ does not exhibit any abrupt nonlinear growth (contrary to the case with $L = 3.8$). Indeed, the non linearly stable
simulation ends up with saturated magnetic islands with structure similar to the last panel shown in Figure 3.
The critical aspect ratio for $L/x_s$ is thus estimated to be $3.75 / 0.8 \simeq 4.7$, that is a value comparable but
slightly lower than the value of $5$ expected for 'sech' profile \citep {jan11}. 

Second, we have investigated the dependence on the equilibrium shear at $x_s$, by varying $a$ between $0.2$ and
$0.4$. The same critical value $L_c \simeq  3.75$ value is recovered whatever the shear value, as illustrated in Figure 6 (left panel)
for $a = 0.2$ case. We can thus conclude to the independence of the critical aspect ratio with the shear for our 'tanh' equilibrium profile.
The reader must note that the minimum length value for linear DTM instability is $L_{min} = 1.26$ and $3.1$ for $a = 0.2$
and $a = 0.4$ respectively. Thus, $L_{min}$ remains lower than $L_c$ for both cases.

Finally, we have investigated the dependence on the distance between the two current sheets by
varying the $x_s$ parameter. The results obtained for a run using $x_s = 0.5$ and $a = 0.2$ are plotted
in right panel of Figure 6. Now, a smaller critical value $L_c  \simeq 2.35$ (as it is in the range between $2.3$ and $2.4$)
can be deduced for this case, leading again to the same critical aspect ratio value of $L_c/x_s = 4.7$. 

We conclude that the critical aspect ratio for the destabilization of the non linear instability is clearly independent of the
details of the equilibrium for the double Harris-like 'tanh' profile. Moreover, it is probably only slightly dependent of the whole profile itself when comparing
the value of order $5$ previously obtained for the 'sech' configuration by \cite {jan11}. This could be due to the presence of the extra magnetic gradient
in the outer region (i.e. for $x \ge x_s$ or $x \le - x_s$) embedding the two current sheets for the 'sech' profile (see Figure 1). Indeed, It has been shown that the presence of external 
current can influence the growth and also saturation level of the magnetic island for the single tearing mode evolution \citep {poy13}.
However, we would draw the attention of the reader to the fact that the difference could also be attributed to the different numerical treatment in the two studies.

By exploring additional cases employing another resistivity value (with $\eta = 3 \times 10^{-4}$), we have checked that the critical aspect ratio
is not resistivity-dependent in agreement with the conclusion previously drawn \citep {jan11}.

\subsection{On the time scale of the explosive reconnection}

Another important question concerns the characteristic time scale for the explosive growth phase. It has been previously shown that it is very weakly dependent or even
independent of the resistivity \citep {wan07, jan11, zha11, akr17}. Moreover, the growth is shown to be faster than a simple exponential growth,
i.e. faster than a scaling law $\sim \exp (\gamma^* t )$ for the perturbed quantities like the kinetic energy, where $\gamma^*$ represents an instantaneous
growth rate. \cite {jan11} suggest a time dependence of the form
$\sim \exp \ [t \ e^{(\gamma^* t)} ]$, and on another hand \cite {akr17} propose another form $\sim \exp ( \gamma^* t^2)$.

In this work, we have examined the time dependence using the kinetic energy variation during the explosive growth, and we have found
that it follows another super-exponential growth. Indeed,
we obtained that, our results can be the best fitted by a double exponential dependence of the form, $E_k \sim \exp [ e^{( \gamma^* (t - t_0) }) ] $, where $t_0$
represents a time instant that can be considered to be close to the onset of the explosive growth phase. 
This is illustrated in Figure 6, for the representative case presented above
in this paper. More precisely, we used $t_0 = 688 \ t_A$ and $\gamma^* = 0.1 \ t_A^{-1}$, in order to correctly approximate the explosive phase in
the range of time values $[688:710]$. For earlier times the non linear instability is not yet triggered, and for later times a saturation ensues due to the final
relaxation (see Figure 2).
Remarkably, this result holds with the same pseudo-rate value of $\gamma^* \simeq 0.1\ t_A^{-1}$ independently of the details of the 'tanh' equilibrium.
Finally, we have also checked that the result is true for different unstable $L > L_c$ values, meaning that it is also independent 
of the 'distance' from the non linear threshold $L - L_c$. This latter point was already visible when comparing the different explosive phases
plotted in Figures 5-6. We complement that, it is obviously different when $L - L_c$ is too high as the Rutherford regime is absent.

One must note that such double exponential growth has been shown to be possible in pure 2D incompressible hydrodynamics due to unstable
configurations driven by the vorticity gradient \citep {den15, kis14}.

\begin{figure}
\centering
 \includegraphics[scale=0.3]{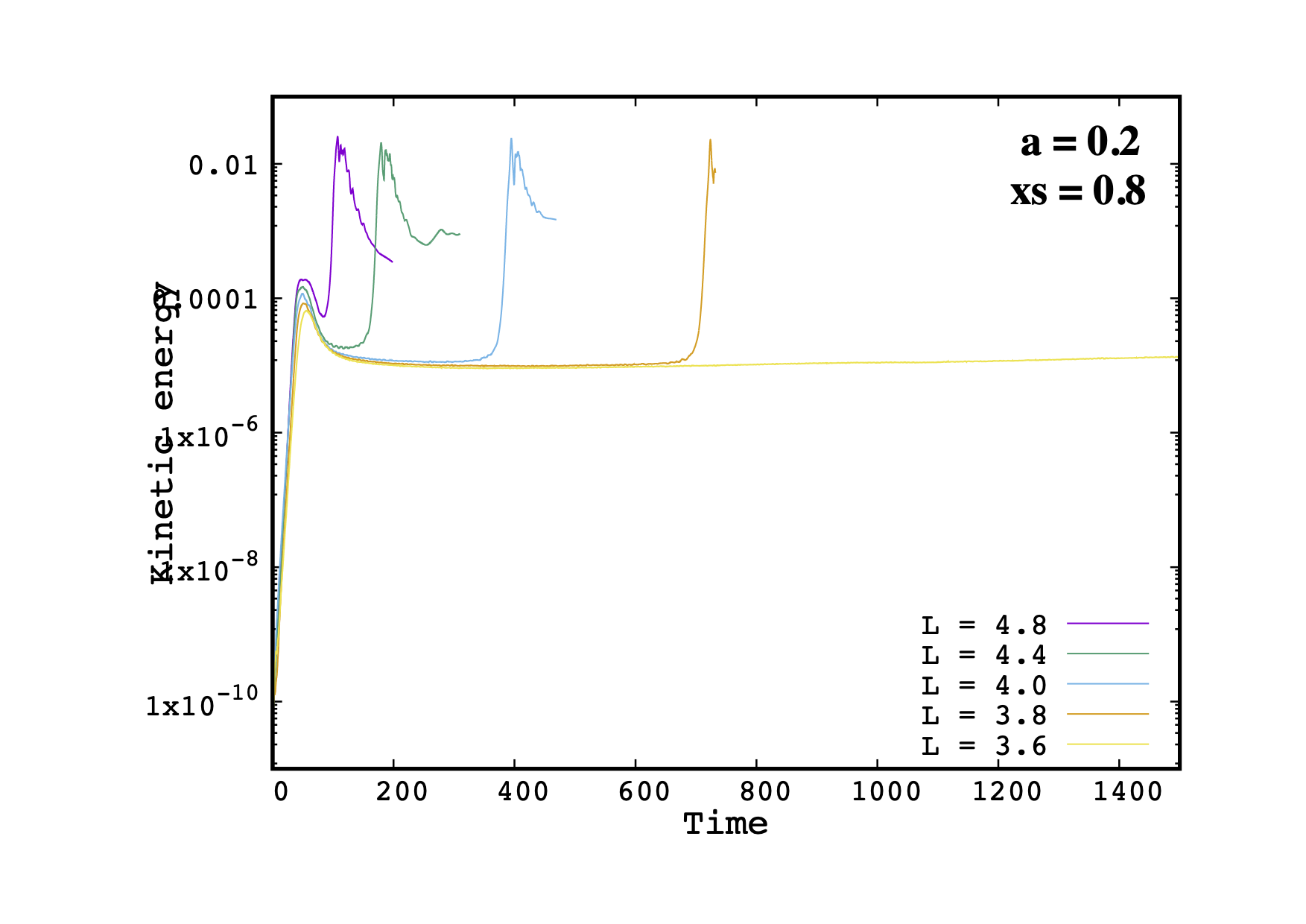}
  \includegraphics[scale=0.3]{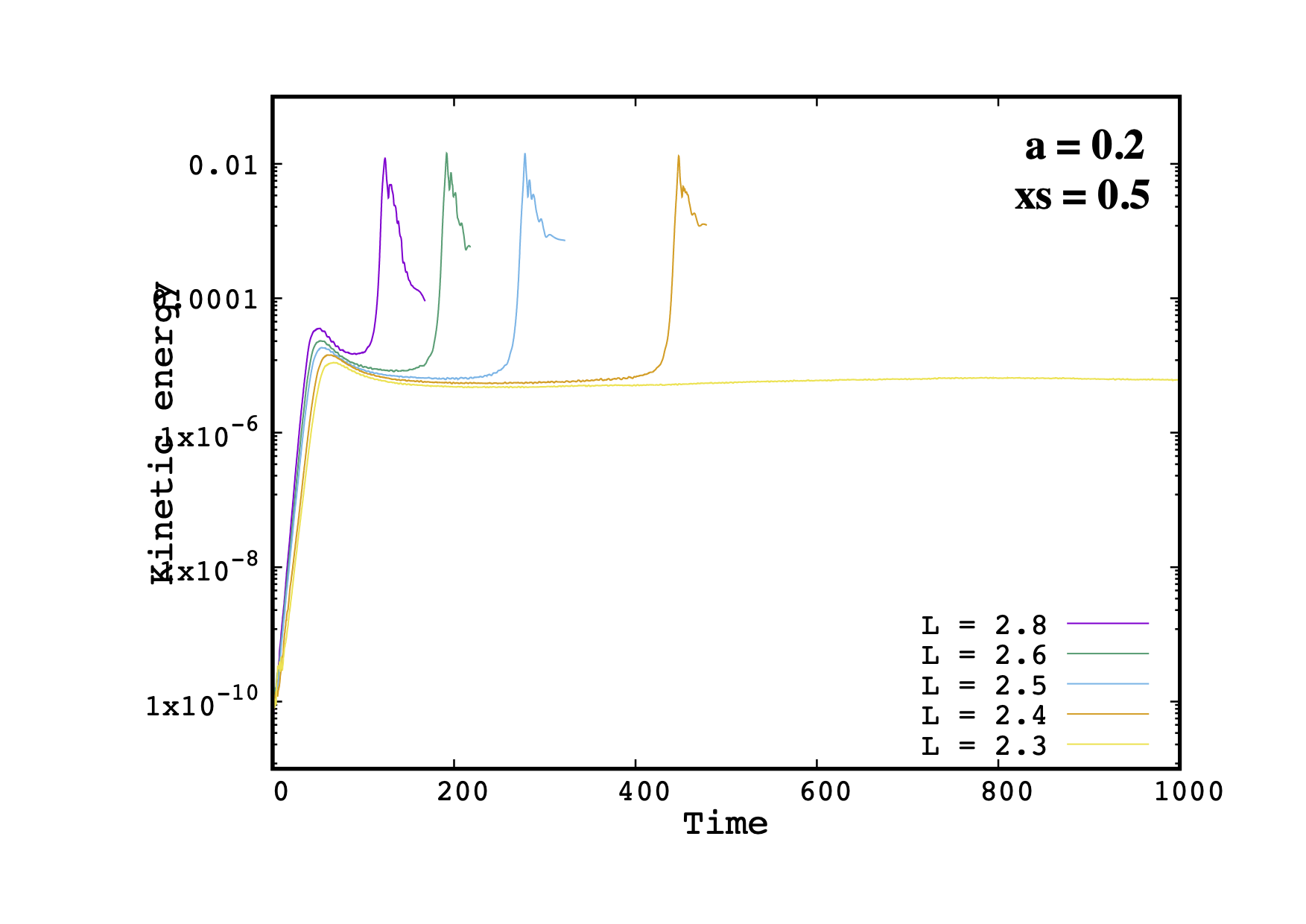}
  \caption{Time evolution of the kinetic energy $E_K$ for cases employing five different length value (see legend). A 'tanh' equilibrium is employed
  with $x_s = 0.8$ and $a = 0.2$, and with $x_s = 0.5$ and $a = 0.2$, in left and right panel respectively.
   }
\label{fig6}
\end{figure}

\begin{figure}
\centering
 \includegraphics[scale=0.5]{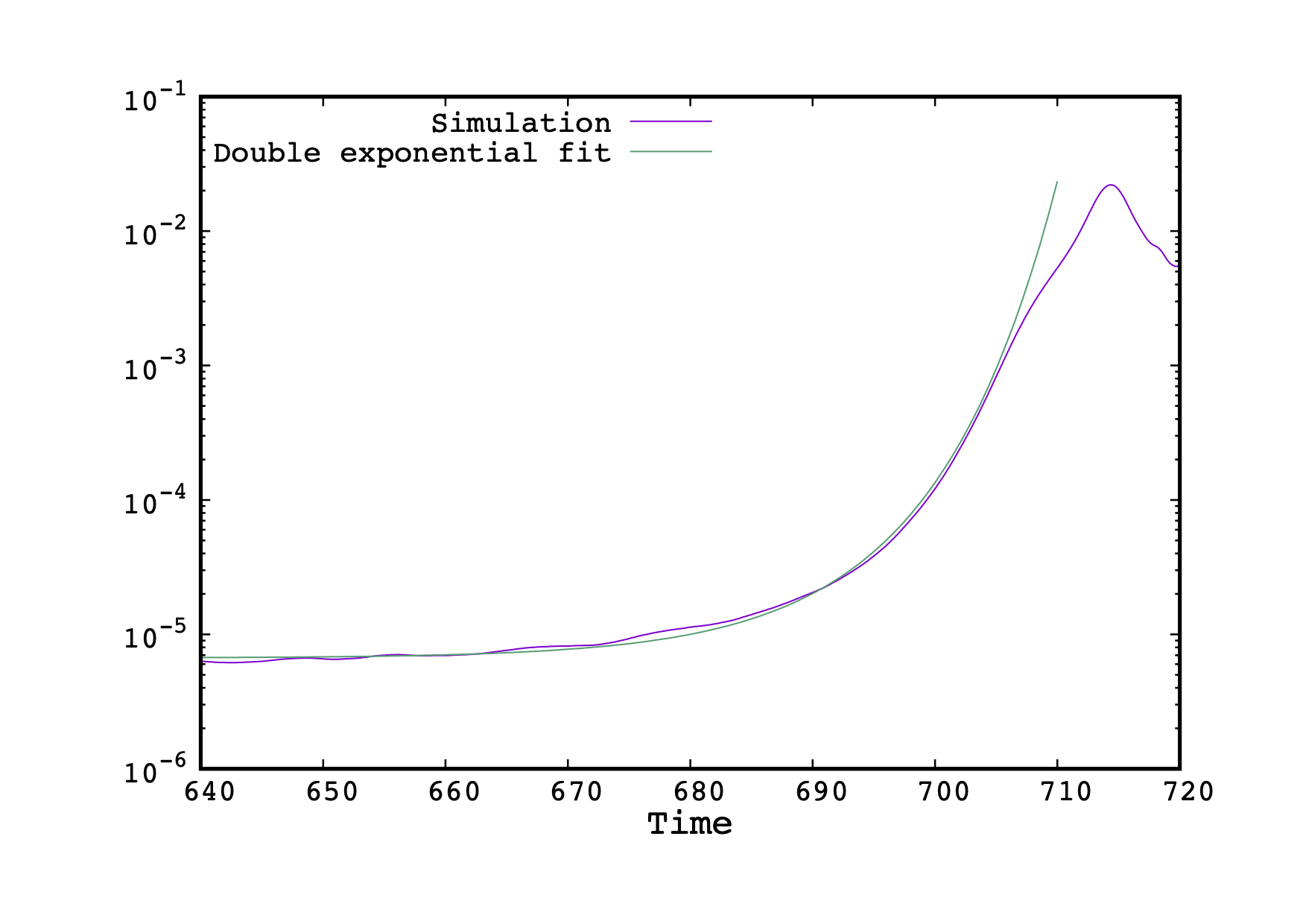}
  \caption{Zoom (centered on the explosive phase) of the time evolution of the kinetic energy for the representative 'tanh' equilibrium case using $a = 0.3$ and
  $x_s = 0.8$. A double exponential fit curve following a law $\sim \exp [ \exp ( \gamma^* (t - t_0))]$ with $t_0 = 688$ and $\gamma^* = 0.1$ is also plotted for comparison.
   }
\label{fig7}
\end{figure}

\section{Conclusion}

In this work, we have confirmed the existence of a critical aspect ratio $L_c/x_s$ for the non linear explosive growth in
double current sheet systems. Its value is independent of the details
of the magnetic equilibrium, and is obtained to be $L/x_s \simeq 4.7$ for the double Harris-like 'tanh' profile considered in this study.
This threshold value is similar and slightly smaller that the one (value close to $5$) deduced by \cite {jan11} on the basis of another MHD equilibrium, namely the 'sech' profile.
The latter small departure could be attributed to the difference in equilibrium current density in the outer region, or to the different numerical
procedure. More work is thus needed in order to further explore this point.

Second, we have examined the time dependence of the explosive phase. As shown in previous studies, we confirm the super-exponential increase
of the kinetic energy $E_K$, as it is faster than a simple exponential law. Moreover, our results suggest that the latter follows a double exponential law
of the form $E_k  \sim \exp [ \exp ( \gamma^* (t - t_0))]$. The value of the characteristic parameter called pseudo-growth rate
is $\gamma^* \simeq 0.1\ t_A^{-1}$, leading to a characteristic time of the order of the Alfv\'en time.

Our results are relevant in the context of plasma environments where multiple current layers can form. This is the case for many astrophysical plasma systems, such as the solar photosphere, solar corona, solar wind, and pulsar wind nebula. In such environment, it is well known that disruptive events leading to magnetic energy release in a sudden way are observed. 
Fast time scale comparable or even smaller than the Alfv\'en one is also required to explain the whole duration of the event.
Thus, the explosive non linear DTM mechanism is a possible route for such disruptions. A second route is provided by the development
of plasmoid chains in the current sheets, when the current layers are able to reach very large aspect ratio (i.e. for very long and/or very thin sheets)
at least $\simeq S_L^{1/3}$, where $S_L = L V_A/ \eta$ \citep {hua17, bat20a, bat20b, bat20c}. This would give aspect ratios higher than $10^3-10^4$ much higher than the values for $L/a$
of the order $10-20$ considered in this work.
In this latter case, an explosive and reconnection mechanism can also be triggered on a time scale that can ever be smaller than the Alfv\'en one
\citep {puc14, com17}. Finally, the two routes are non exclusive, as the two mechanisms can also be at work in a simultaneous way
\citep {bat17}. Anyway, a more realistic study would require to include the initial process of formation of the two current sheets, that are
assumed to be preformed in the present work.

\


\begin{thebibliography}{}


\bibitem[Akramov \& Baty(2017)]{akr17} Akramov, T., \& Baty, H. 2017, PoP, 24, 082116,
https://doi.org/10.1063/1.5000273

\bibitem[Baty(2017)]{bat17} Baty, H. 2017, ApJ, 837, 74,
https://doi.org/10.3847/1538-4357/aa60bd

\bibitem[Baty(2019)]{bat19} Baty, H. 2019, ApJSS, 243, 23,
https://doi.org/10.3847/1538-4365/ab2cd2

\bibitem[Baty(2020a)]{bat20a} Baty, H. 2020a, arXiv:2001.07036
https://ui.adsabs.harvard.edu/abs/2020arXiv200107036B

\bibitem[Baty(2020b)]{bat20b} Baty, H. 2020b,  arXiv:2003.08660
https://ui.adsabs.harvard.edu/abs/2020arXiv200308660B

\bibitem[Baty(2020c)]{bat20c} Baty, H. 2020c,  arXiv:2006.15013
https://ui.adsabs.harvard.edu/abs/2020arXiv200615013B

\bibitem[Biskamp(2009)] {bisk09} Biskamp, D. 2009,
Nonlinear Magnetohydrodynamics, (Cambridge University Press).
https://doi.org/10.1017/CBO9780511599965

\bibitem[Comisso et al.(2017)]{com17} Comisso, L., Lingam, L., Huang, Y.~M., \& Bhattacharjee, A. 2017, ApJ, 850, 142,
https://doi.org/10.3847/1538-4357/aa9789

\bibitem[Denisov(2015)]{den15} Denisov, S. A. 2015, Proc. Amer. Math. Soc. 143, 1199-1210,
https://arxiv.org/pdf/1201.1771.pdf
  
\bibitem[Huang et al.(2017)]{hua17} Huang, Y.~M., Comisso, L., \& Bhattacharjee, A. 2017, ApJ, 849, 75,
https://doi.org/10.3847/1538-4357/aa906d

\bibitem[Ishii et al.(2002)]{ish02} Ishii, Y., Azumi, M., \& Kishimoto, Y. 2002, PRL, 89, 205002,
https://doi.org/10.1103/PhysRevLett.89.205002

\bibitem[Kiselev \& Sverak(2014)]{kis14} Kiselev, A.,  \& Sverak, V. 2014, Annals of Mathematics
Vol. 180, No. 3, 1205-1220,
 https://arxiv.org/pdf/1310.4799.pdf

\bibitem[Janvier et al.(2011)]{jan11} Janvier, M., Kishimoto, M. Y., \& Li, J. Q. 2011, PRL, 107, 195001,
https://doi.org/10.1103/PhysRevLett.107.195001

\bibitem[Priest \& Forbes(2000)]{pri00} Priest, E.~R., \& Forbes, T.~G. 2000, Magnetic Reconnection
(Cambridge: Cambridge Univ. Press),
https://doi.org/10.1017/CBO9780511525087

\bibitem[Poyé et al.(2013)]{poy13} Poyé., A., Agullo, O., Smolyakov, A., Benkadda, S., \& Garbet, X. 2013, PoP, 20, 020702,
https://doi.org/10.1063/1.4791653

\bibitem[Pucci \& Velli(2014)]{puc14} Pucci, F., \& Velli, M. 2014, ApJL, 780, L19,
https://doi.org/10.1088/2041-8205/780/2/L19

\bibitem[Wang et al.(2007)]{wan07} Wang, Z. X., Dong, J. Q., Lei, Y. A., Long, Y. X., Mou, Z. Z.,  \& Qu, W. X. 2007, PRL, 99, 185004,
https://doi.org/10.1103/PhysRevLett.99.185004

\bibitem[Wei et al.(2020)]{wei20} Wei, L., Yu, F., Ren, H. J., \& Wang, Z. X. 2010, AIP Advances, 10, 055111,
https://doi.org/10.1063/5.0007522

\bibitem[Zhang \& Ma(2011)]{zha11} Zhang, C. L., \& Ma, Z. W. 2011, PoP, 18, 052303,
https://doi.org/10.1063/1.3581064


\end{thebibliography}
\end{document}